Efficient absolute interface energy calculations for heterostructures: Synergy between localized basis sets and surface passivation techniques

# Efficient absolute interface energy calculations for heterostructures: Synergy between localized basis sets and surface passivation techniques

Sreejith PALLIKKARA CHANDRASEKHARAN, Sofia APERGI, Charles CORNET*, Laurent PEDESSEAU*

*Univ Rennes, INSA Rennes, CNRS, Institut FOTON – UMR 6082, F-35000 Rennes, France*

*Corresponding Authors

Correspondence to *Charles Cornet* (*charles.cornet@insa-rennes.fr*) *and Laurent Pedesseau* (*laurent.pedesseau@insa-rennes.fr*)

## Abstract

Heterostructures combining diverse physico-chemical properties are increasingly in demand for a wide range of applications in modern science and technology. However, despite their importance in materials science, accurately determining absolute interface energies remains a major challenge. Here, we present a computationally efficient framework for determining interface energies by incorporating a surface-passivation technique, demonstrated using pseudo-H* passivation with a localized-basis-set method and an explicit chemical potential. This framework is applied to calculate absolute interface energies and analyze the electronic properties of quasi-lattice-matched and lattice-mismatched III-V/Si interfaces, with results compared to conventional reconstructed-surface calculations. By combining localized basis sets with surface passivation techniques, this framework allows for accurate estimation of absolute interface energies in heterogeneous material systems. This approach effectively addresses issues associated with surface reconstructions while significantly reducing computational costs within the framework of density functional theory, and moreover offers considerable potential for calculating interface energies across diverse material systems.

## Introduction

As H. Kroemer emphasized in his Nobel lecture, a deep understanding of interfaces between dissimilar materials is crucial, as they play a pivotal role in determining device performances[1,2]. Especially, the knowledge of interface energy, whose atomic structure is very hard to resolve experimentally, provides a fundamental framework for predicting interface configuration, its physical properties, and may help to clarify processes involved during heteroepitaxy, heterointegration, or chemical synthesis on a substrate. Notably, the chemical mismatch between substrate atoms and the deposited layer may lead to the formation of a hybrid interface[3]. Evaluating the energetics of heterointerfaces offers valuable insights into thermodynamic stability[4-7], interfacial bonding[6-9], electronic properties[10-17], strain relaxation processes[11,18,19], and defect generation[20], particularly in polar-on-non polar systems[21-23]. Such insights are essential in optimizing growth conditions[24,25], reducing defects[22,26], and enhancing the quality and performance of heterostructure devices for the next-generation technologies[27-29].

Numerous studies have investigated interface energetics using a variety of experimental techniques. These include direct methods such as scanning microscopes[30], photoelectron spectroscopy[31], calorimetric measurements[32], as well as indirect approaches like inverse Wulff-Kaischew constructions and relative energy assessments[33]. Despite these significant advances, the precise determination of absolute interface energy values remains largely underexplored, mainly inherent to challenges associated with obtaining accurate measurements as well as realistic and accurate theoretical simulations.

*Ab initio* atomistic studies have significantly contributed to our understanding of the atomic arrangements at the heterointerfaces[34]. A broad survey of the literature shows that various groups worldwide have addressed the challenge of determining the interface energy calculations using different strategies. Several methods have been developed to evaluate interface energies in heteroepitaxial systems. These approaches can be classified into two families: the relative energy methods (also known as the “superlattice" methods) and the absolute energy methods. The relative energy methods provide only relative values of interface energies and require a convergence study of the total energy as a function of the material thickness[7,13,14,35-37]. Although this approach enables the establishment of stability rankings among different structures, it does not provide quantitatively absolute interface energies. An important contribution within this approach was made by W. Zhang and co-workers[38], who examined the effect of chemical potential variations on interface energy calculations. More recently, T. Hannappel and co-workers emphasized the importance of incorporating chemical potential into interface energy calculations within a thermodynamic framework for GaP/Si[7,13,14]. By applying the relative energy method, they clarified charge compensation mechanisms and compared interfaces stability over the accessible thermodynamic range. A similar perspective was also adopted by S-H Wei and co-workers in their work on II-VI/III-V interfaces, discussing different charge compensated interfaces[39].

In contrast, calculations aiming at determining the absolute interface energy offer a quantitative measure of interface stability and provide fundamental insights essential for understanding the behavior and properties of heterointerfaces. This approach, often viewed as computationally demanding, typically requires constructing a slab model that includes a bottom surface, the bulk substrate, the hybrid interface region, an epitaxial overlayer, and a top surface. The interface energy is then determined by systematically subtracting the contributions of surface and bulk energies from the total energy of the slab. Nevertheless, many studies have overlooked the broader influence of surface effects on the accuracy of interface energy calculations. In particular, the treatment of surface dipoles, slab thickness, and vacuum spacing are critical factors that can significantly affect the computed energetics. Addressing all these issues is crucial for obtaining reliable results. Some investigations have attempted to account for this by explicitly considering reconstructed surfaces of various degrees of complexity[23]. These challenges of surface reconstructions can be avoided by considering use of surface passivation techniques on the surfaces to mimic the bulk. In fact, surface passivation techniques are diverse (H passivation, functional group such as OH or $CH_3$ passivation, pseudo-hydrogen (pseudo-H* or H*) passivation, pseudo atomic charge passivation, semiempirical electrostatic corrections). Among these, one of the most common and widely applied approaches involves the use of pseudo-H* passivation[40-44,23]. Passivation of surface slabs using pseudo atomic charges has also been explored for the ferroelectric material $BaTiO_3$. In this approach, pseudo charges on oxygen atoms were introduced to artificially compensate for the internal charge within the supercell slab and to cancel the spurious electric dipole artifact in the calculations, as demonstrated by J. Junquera and co-workers[45]. In recent works, N. Marom and co-workers[46] employ pseudo-H* passivation to study the interface energies; however, their framework did not incorporate chemical potentials, which in fact play a crucial role in the thermodynamic description of interface energetics, and replies on a plane-wave code, which handles large vacuum regions less efficiently than localized basis set codes. Moreover, J. Zhu and co-workers[47] advanced this direction by combining surface passivation schemes with chemical potential variations, thereby achieving a more comprehensive thermodynamic representation of interface energies. Nonetheless, the use of a limited vacuum thickness of about 15 Å could introduce finite-size effects, and electrostatic artifacts a constraint arising from the high computational cost of plane-wave-based calculations.

Despite these advances, none of the aforementioned plane-wave-based studies successfully captured surface–surface dipole interactions, as their accurate treatment requires sufficiently large vacuum separations (at least 150-250Å depending on interactions) to avoid spurious electrostatic coupling between periodic images, which is computationally expensive in plane-wave frameworks. To address this limitation, one effective strategy to mitigate surface-related artifacts involves combining the use of localized basis set codes such as SIESTA[48-50] or FHI-aims[51] alongside surface passivation techniques, including variation of the chemical potential. Specifically, localized basis sets such as numerical atomic orbitals (NAOs) offer a more efficient alternative. In particular, the SIESTA code, which employs NAOs, has been used to calculate absolute interface energies for systems including reconstructed surfaces, despite the associated high computational costs[22,23]. In those studies, both the bottom and top surfaces were atomically reconstructed based on assumed configurations extracted from the literature reports. The calculated interface energies are expressed as a function of chemical potential variations ($\Delta\mu$) within the thermodynamic regime. A major challenge in calculating absolute interface energy lies in constructing a supercell that accommodates various surface reconstructions, while simultaneously: i) minimizing electric dipole interactions between the supercell and its periodic images by introducing large vacuum region; ii) achieving convergence of the absolute interface energy by incorporating a substantial amount of material on both sides of the interface; and iii) exploiting symmetry as much as possible to reduce computational cost.

In this work, we combine localized basis sets calculations with surface passivation techniques to present a robust and versatile approach for calculating absolute interface energies as a function of chemical potential dependencies, enabling its application across a broad spectrum of heteroepitaxial systems. This method significantly enhances slab symmetry. Furthermore, it fundamentally addresses challenges inherent to reconstructed surfaces in density functional theory (DFT) calculations[52,53], including errors associated with dipole interactions originating from both the top and bottom surfaces of the heterostructure, as well as from the interface itself. Furthermore, previous studies often overlook the effects of electric dipoles caused by insufficient vacuum spacing and the critical role of chemical potential. Although some DFT codes incorporate dipole-dipole corrections to partially address multipole interactions, these corrections are not sufficient and sometimes inadequate. This study reveals that the combined approach of localized basis sets and passivated surfaces (CLAPS) for interface energy calculations proposed here provide a robust and accurate means for determining the absolute interface energies of various heterostructures, including quasi-lattice-matched (GaP/Si) or strained ones (GaAs/Si). Finally, we demonstrate how the combined localized basis sets-surface passivation technique sweeps away errors resulting from electric dipole effects at the hetero-interface. This is achieved through comprehensive non-stoichiometric abrupt interface calculations, supported by systematic investigations of structural thickness, vacuum size, bond lengths, and charge densities, along with analyses of electrostatic potentials and electronic band structures. In addition, we compare the computational expenses and environmental sustainability of both approaches.

## Results

For an accurate computational determination of the absolute interface energy between two dissimilar materials, the influence of the top and bottom surfaces should be carefully cancelled. Assuming a given atomic reconstruction for these surfaces thus leads to specific dipoles or surface-stress-related effects, which can ultimately influence the accuracy of the calculated interface energy. For the surface passivation technique, we employed the pseudo-H* passivation, which replaces surface reconstruction atoms with pseudo atoms bonded directly to the bulk. The synergy between local basis set and a passivation technique offers three main advantages. First, top and bottom surfaces passivation allows the top and bottom surfaces of the slab to mimic the bulk properties of a heterostructure. Second, the impact of multipoles disappears dramatically and naturally, a result that is beyond the capabilities of traditional DFT all-electron or plane-wave codes. Third, when combined with localized basis set codes such as SIESTA, this approach ensures convergence of the electric dipole and total energy for a polar slab, with negligible contributions from Coulombic interactions and surface dipoles arising from slab images[54]. Recently, D. Bennett *et al.*[55] demonstrated that a localized basis offers superior scaling with system size and greater suitability to low-dimensional system, advantages that we exploit in this study.

Here, absolute interface energies for a Ga-abrupt GaP/Si(001) (quasi-lattice-matched) interface (where Si atoms are only bonded to Ga atoms at the interface) are compared using the H* passivation and the reconstructed surface approaches. The approach is then applied to the bi-axially strained GaAs/Si(001) case and discussed with respect to the quantum-size effects of slab thickness on the non-converging behavior in interface energy calculations.

## Absolute interface energy: reconstructed surfaces

The absolute interface energy of different III-V/Si systems was first calculated using stable, reconstructed top and bottom surface slabs. This approach enables the calculation of the absolute interface energy by eliminating the bulk and surface energy contributions from the total energy of the interface supercell slab per unit area (details in methods section). A thorough understanding of the predominant surface reconstructions of the two different materials in the heterostructure is crucial for this approach. The analysis of surface reconstructions must account for the system's polarity, the influence of electric dipoles on DFT calculations, and the associated computational costs. A comprehensive study of the absolute interface energies of III-V/Si heterostructures, based on free surfaces models, has been presented in previous studies[22,23]. In the present work, we employed a slab model, illustrated in Fig. 1a-d. Figure 1a shows a schematic of a supercell representing an abrupt GaP/Si(001) interface. This model includes stable reconstructed GaP(001)md(2x4) surface on the top (Fig. 1b), and a reconstructed Si(001) surface on the bottom (Fig. 1c). A Ga-abrupt interface was considered, with Ga atoms directly bonded to the silicon ones, as shown in Fig. 1d. To suppress electrostatic charge-charge and multipole interactions between the slab and its periodic images, a vacuum region of approximately 400 Å was introduced, following the approach detailed in the ref.[23], and further described in methods section. Such vacuum thickness could not reasonably be implemented with plane-waves approaches.

### Absolute interface energy: H* passivation on surfaces

The second modeling strategy proposed here is to use a passivation technique such as pseudo-H* charge for surface modeling to overcome the challenges associated with limited knowledge of surface reconstructions across various heterostructures, the large system sizes needed to capture these effects, and the influence of electric dipole interactions. The passivation technique effectively compensates the dangling bonds at the surface, thereby mimicking the bulk-like electronic environment[41]. Both approaches require separate surface energy contributions and therefore rely on independent surface slab calculations. In this study, the pseudo-H* approach accounts for these contributions by treating surface slabs passivated on both sides independently, thereby incorporating the energies of surface atoms and pseudo-H* passivation without the need for explicit surface energy calculations. This treatment simplifies calculations by avoiding a separate determination of pseudo-H* energies while consistently including surface atom contributions. Similarly, the interface energy is defined as the excess or deficit of the total system energy relative to the combined energies of the top and bottom surface slabs and the bulk regions. To model surface passivation, we employed pseudo-H* atoms based on the principles of $sp^3$ hybridization and the local tetrahedral symmetry of surface bonds. The valence charge of each pseudo atom was adjusted to ensure overall charge neutrality, with net charges assigned as follows: 0.75 e for pseudo atoms bonded to group V atoms, 1.25 e for those bonded to group III atoms, and 1.0 e for those bonded to Si atoms. In our previous works, we have validated the reliability of this methodology for calculating both polar and non-polar surface energies[23].

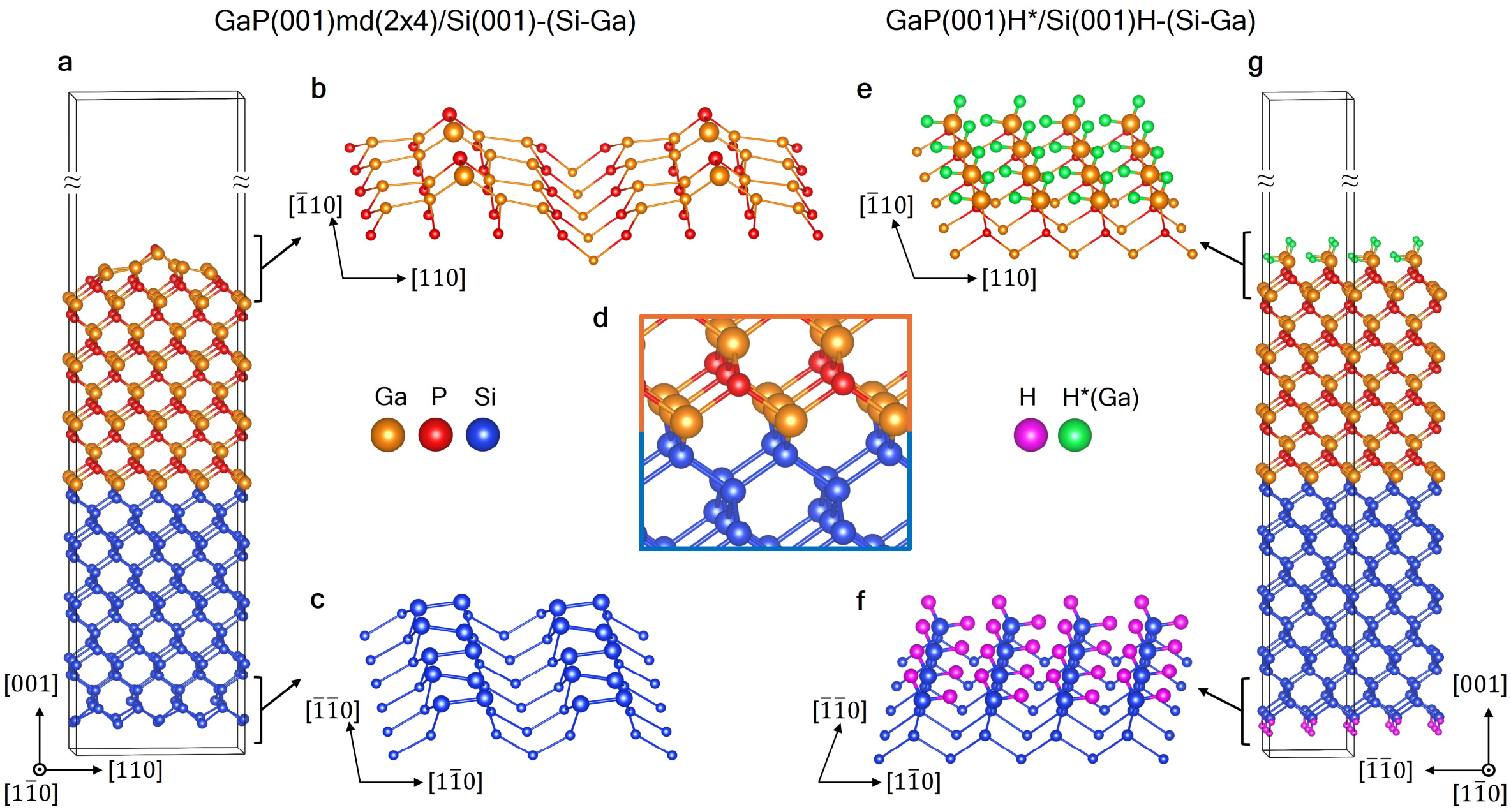

**Fig. 1. Representation of relaxed slab models considered for determining absolute interface energies of GaP/Si.** **a** Total slab model used for implementation of the reconstructed surface method, with **b** the reconstructed GaP(001)md(2x4) on top and **c** the Si(001) bottom surface for the **d** abrupt interface configuration. **e** top and **f** bottom view of H* and H passivated surface terminations to compensate charges, and **g** the overall supercell model to determine absolute interface energy with the H* passivation method.

In general, the absolute interface energy ${}^{\mathrm{I}}\gamma^{\mathrm{Top}}_{\mathrm{Bottom}}$ for interface $\mathrm{I}$, where $\mathrm{Top}$ and $\mathrm{Bottom}$ represent the specific surface slabs of the material system is defined as:

$$ {}^{\mathrm{I}}\gamma^{\mathrm{Top}}_{\mathrm{Bottom}} = \frac{E^{\mathrm{int}}_{\mathrm{slab}} - \sum_{i=Top,Bottom}\left(\frac{1}{2}E^{\mathrm{surf_slab}}_{\mathrm{i}}\right) - \sum_{\mathrm{j}}\left(N^{\mathrm{bulk}}_{\mathrm{j}}\mu^{\mathrm{bulk}}_{\mathrm{j}}\right) - \sum_{\mathrm{j,k}}\left(\Delta N_{\mathrm{j}}\ \mu_{\mathrm{k}}\right)}{A} \qquad (1) $$

where $E^{\mathrm{int}}_{\mathrm{slab}}$ is the total energy of the interface supercell slab, $E^{\mathrm{surf_slab}}_{\mathrm{i}}$ is the $\mathrm{Top}$ and $\mathrm{Bottom}$ surface slabs energies, $\mu^{\mathrm{bulk}}_{\mathrm{j}}$ is the chemical potential of the $N^{\mathrm{bulk}}_{\mathrm{j}}$ atoms, $\Delta N_{\mathrm{j}}$ is the stoichiometry and $\mu_{\mathrm{k}}$ is the chemical potential of the elements k of the composing material j. A is the in-plane surface area of the slabs.

Here, the analysis contains three distinct elements, with group-III and V atoms at the top and Si at the bottom of the heterostructure. As discussed in our previous work[23], the total slab, surface slabs, and interface each maintain distinct stoichiometries. The stoichiometry is quantified by $\Delta N_{III-V}$ for the III-V material and $\Delta N_{Si}$ for silicon.

From Eqn. (1), final expression for the absolute interface energy in the studied heterostructure is:

$$ {}^{I}\gamma_{Si}^{III-V} = \frac{E_{slab}^{int} - \frac{1}{2}\left(E_{III-V}^{surf_slab}\right) - \frac{1}{2}\left(E_{Si}^{surf_slab}\right) - N_{III-V}^{bulk}\mu_{III-V}^{bulk} - N_{Si}^{bulk}\mu_{Si}^{bulk} - \Delta N_{III-V}\mu_{V} - \Delta N_{Si}\mu_{Si}}{A} \tag{2} $$

where $E_{slab}^{int}$ is the total energy of the interface supercell slab, $E_{III-V}^{sub-surf}$ and $E_{Si}^{sub-surf}$ are the total energies of III-V and Si surface slabs. The chemical potentials of bulk III-V and Si atoms are denoted as $\mu_{III-V}^{bulk}$ and $\mu_{Si}^{bulk}$, while $N_{III-V}^{bulk}$ and $N_{Si}^{bulk}$ are the corresponding number of bulk atoms. The stoichiometries of the III-V and Si materials are given by $\Delta N_{III-V}$ and $\Delta N_{Si}$ with $\mu_{V}$ and $\mu_{Si}$ representing the chemical potentials of top (V) and bottom (Si) species, and A is the in-plane surface area of the slabs. The relationship simplifies naturally under the silicon surface condition ($\Delta N_{Si} = 0$). Interestingly, pseudo-H* energy is embedded in the surface slab term, eliminating the need for separate accounting.

The schematic representation of the relaxed slab considered for H* passivation methodology is illustrated in Fig. 1d-g. As shown in Fig. 1d, similar Ga-abrupt interfaces are modeled, for both approaches studied. Here, the atoms of the top surface (Fig. 1e) are passivated with H*(Ga) and a compensation charge of 1.25e, while the atoms of the bottom Si surface (Fig. 1f) are passivated with H atoms to neutralize charges. The complete H* passivated supercell used for absolute interface energy calculations is depicted in Fig. 1g. Notably, the supercell is reduced to half the size of that used in the reconstructed surface method (Fig. 1a), offering computational efficiency. Additionally, a vacuum thickness of approximately 400 Å is introduced to eliminate spurious electrostatic charge-charge and dipole-dipole interactions between the slab and its image (details in methods section).

### GaP/Si: quasi-lattice-matched interface

Absolute interface energies with both reconstructed surfaces and H* passivated surfaces approaches are computed within the thermodynamic stability range for a quasi-lattice matched GaP/Si heterointerface. For GaP, we used the thermodynamic range determined by the heat of formation energy of -0.928 eV[23], so that the interface energy is given between the two thermodynamic limits corresponding to the formation of bulk Ga and P respectively: the Ga-rich and P-rich limits. Here, we have used the previously reported Ga-abrupt absolute interface energies for two possible surface reconstructions of GaP in the (001) direction, specifically GaP(001)(2x4) and GaP(001)md(2x4)[23]. Interface energies were calculated as a function of phosphorus chemical potential variations ($\Delta\mu_P$)[22,23], and the resulting trends are illustrated in Fig. 2a, with the corresponding values provided in the Supplementary information (see Fig. S1 and Table S1). From our previous results, the reconstructed surfaces method led to interface energies of 40.8 meV/Å² (Ga-rich) and 72.0 meV/Å² (P-rich) for GaP(001)(2x4), and 38.5 meV/Å² (Ga-rich) and 69.7 meV/Å² (P-rich) for GaP(001)md(2x4), respectively. On the other hand, using the H* passivation approach, we obtained values of 43.6 meV/Å² under Ga-rich conditions and 74.7 meV/Å² under P-rich conditions. These results align well with those derived from the surface reconstruction method, having an error margin of less than 5 meV/Å² (attributed to dipole interactions, as discussed later).

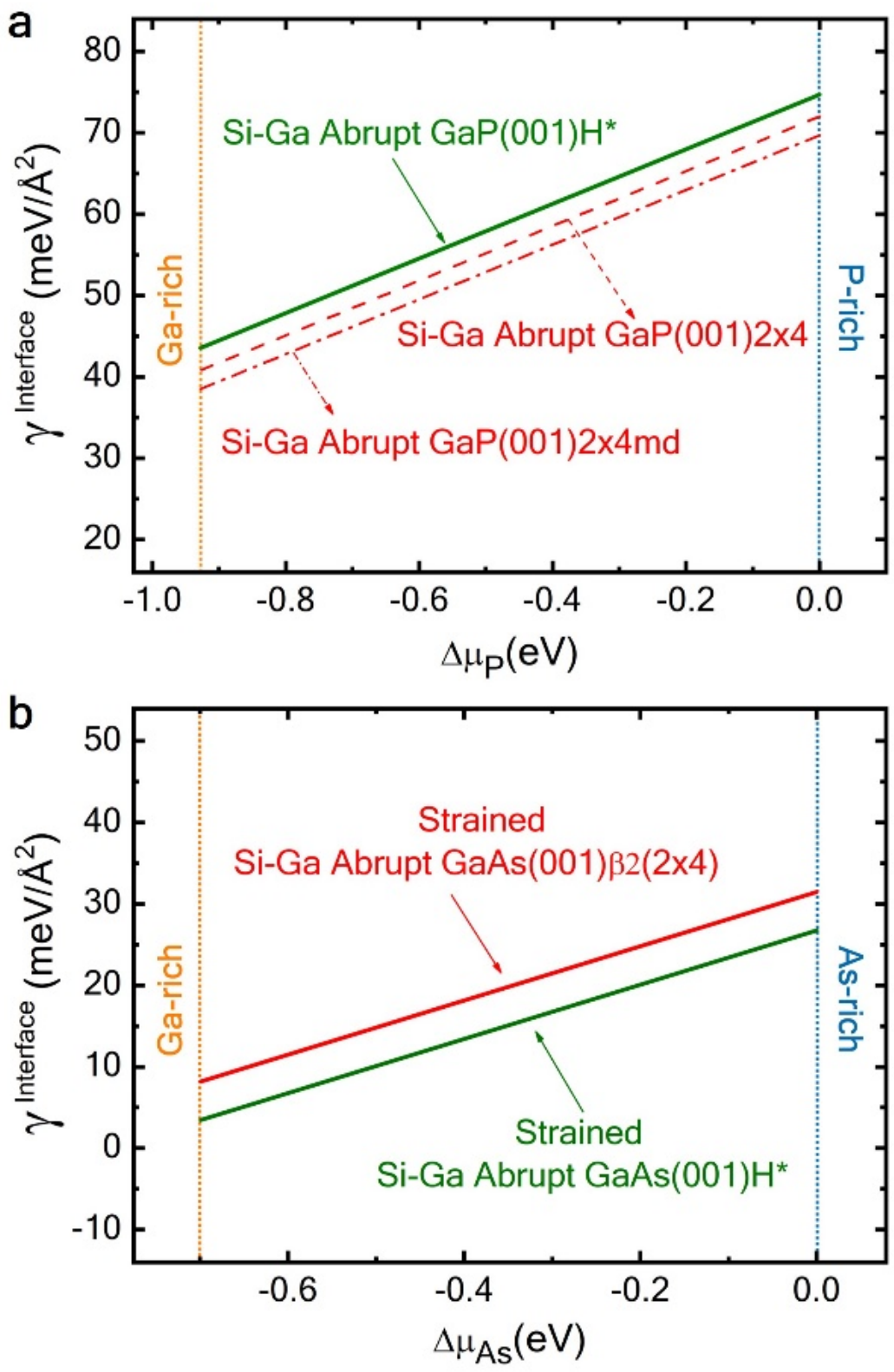

**Fig. 2. Comparison of absolute interface energies ($\gamma^{\mathrm{Interface}}$) calculation following the two different methods. a** Absolute interface energies for GaP/Si with Ga-abrupt interfaces were calculated using the H* passivated method (green) and compared with previously reported GaP(001)(2x4) and GaP(001)md(2x4) surface reconstruction methods (red), as a function of the chemical potential variation $\Delta\mu_{\mathrm{P}}$. **b** Absolute interface energies for strained GaAs/Si with Ga-abrupt interfaces, calculated using the surface reconstruction method (red) and compared to the H* passivation approach (green), as a function of the chemical potential variations $\Delta\mu_{\mathrm{As}}$.

### GaAs/Si: lattice mismatched & strained growth

This study is now generalized to a lattice-mismatched heterointerface. In the following section, we examine the biaxially strained epitaxial integration of GaAs on Si using both methodologies. For GaAs, the surface and interface energies are governed by the variations of the chemical potentials of arsenic $\mu_{\mathrm{As}}$ and gallium $\mu_{\mathrm{Ga}}$, which reflect the thermodynamic conditions under which each element can exist within the bulk or at the surface of the GaAs material. Indeed, these chemical potentials play a critical role in determining the stability and morphology of the heterointerface. In this work, the heat of formation of GaAs, $\Delta\mathrm{H_f}$(GaAs), was determined to be -0.70 eV, consistent with previously reported values in the literature[56].

A comprehensive description of the methodology is provided in the methods section. This value delimits the thermodynamic stability window of GaAs, extending from Ga-rich to As-rich conditions, as illustrated in Fig. 2b.

In the early stages of epitaxial growth, before the plastic relaxation, GaAs grows pseudomorphically on the Si substrate, accommodating the lattice mismatch (experimentally measured ≈ 4%)[57,58]. This results in a bi-axial compressive strain in the plane of the interface and in a corresponding elongation along the (001) growth direction. To accurately account for this strain, we considered both the bi-axially strained bulk and surface structures, as outlined in the Supplementary information (see details in Supplementary Fig. S2 and Supplementary note 1).

To determine the interface energies of a strained GaAs/Si heterostructure using a reconstructed approach, we analyzed the strained GaAs(001)β2(2x4) surface on top, one of the stable GaAs surfaces in the (001) direction[56], and stable reconstructed Si(001)c(4x2) at the bottom[59]. Supplementary Fig. S3 illustrates the studied GaAs surfaces, with further details provided in Supplementary note 1. The methodology for determining absolute interface energies is comprehensively described in the methods section. A schematic representation of the reconstructed GaAs/Si interface is illustrated in Supplementary Fig. S4a-d. The interface energy varies from 8.2 meV/Å² (Ga-rich) to 31.5 meV/Å² (As-rich), within the thermodynamic range, as shown in Fig. 2b, and summarized in Supplementary Table S1.

Next, the strained interface energy is calculated using the pseudo-H* passivation on the surfaces to ensure charge compensation and to mimic bulk-like properties on the surfaces, as previously demonstrated for the GaP/Si heterostructure. Supplementary Fig. S3 shows a schematic comparison of the strained GaAs surface slab structures for the GaAs/Si interface using surface-reconstructed and passivated surface technique, with the supercell size reduced to one-fourth. Additionally, Supplementary Fig. S4 further compares the GaAs/Si interface slabs for both approaches, incorporating an additional supercell size reduction by half. Using this approach, the calculated interface energies are 3.5 meV/Å² and 26.7 meV/Å² under extreme Ga-rich and As-rich thermodynamic conditions, respectively.

Figure 2b compares the calculated interface energies from the H* passivation approach with those from the reconstructed across the thermodynamic range, and Supplementary Table S1 summarizes the results. Again, these results highlight the robustness of the CLAPS approach for determining absolute interface energies, achieving an accuracy within 5 meV/Å². Importantly, CLAPS requires a significantly smaller lateral supercell size compared to conventional calculations involving reconstructed surfaces. Compared to the GaP/Si interface, the GaAs/Si(001) heterostructure exhibits similar thermodynamic behavior, showing lower interface energies toward the Ga-rich side due to the similar Ga-abrupt configuration. Notably, under Ga-rich conditions, the bi-axially strained GaAs/Si interface energies are drastically lowered relative to GaP/Si. The observed trend in the interface energies for bi-axially strained heterostructures can be attributed to a combined effect of chemical inhomogeneity at the interface and structural distortions resulting from lattice misfit[18]. Recent calorimetric measurements by Calvin *et al.* even reported negative interface energies in the ZnS/InP heterostructure, which exhibits a pronounced lattice mismatch of approximatively 8.4%[32].

The evolution of the calculated interface energy with the GaAs thickness considered in the slab is then studied by computing different slabs having different thicknesses (Fig. 3a and Fig. 3b). Both approaches exhibit convergence in interface energy per atom (N(Ga, As)) with increasing GaAs thickness under Ga-rich conditions (Fig. 3a), adjusted with solid lines as a guide to the eyes (see Supplementary note 2), where larger discrepancies between the two methods are observed at lower thickness likely due to surface dipole effects (discussed later in more details). Additionally, a linear divergence (not shown here) in interface energy values emerges with increasing GaAs slab thickness due to underestimated bulk energy contributions. To address this issue, we applied the Boettger correction[60,61]. The resulting corrected interface energies are shown in Fig. 3a (inset), and exhibit no further evolution with the GaAs thickness, demonstrating the reliability of this correction even for large supercells and enabling accurate comparison between the two approaches. However, it is important to note that this correction does not account for surface dipole contributions. In contrast, the H* passivation method effectively eliminates the influence of dipoles arising from surface reconstructions, leading to a more accurate interface energy compared to the reconstructed surface method. In addition, we analyzed the interface distortion relative to the bulk and found that local strain shows no significant difference between the two approaches, whereas passivation mechanism further reduces these distortions (see Supplementary note 3).

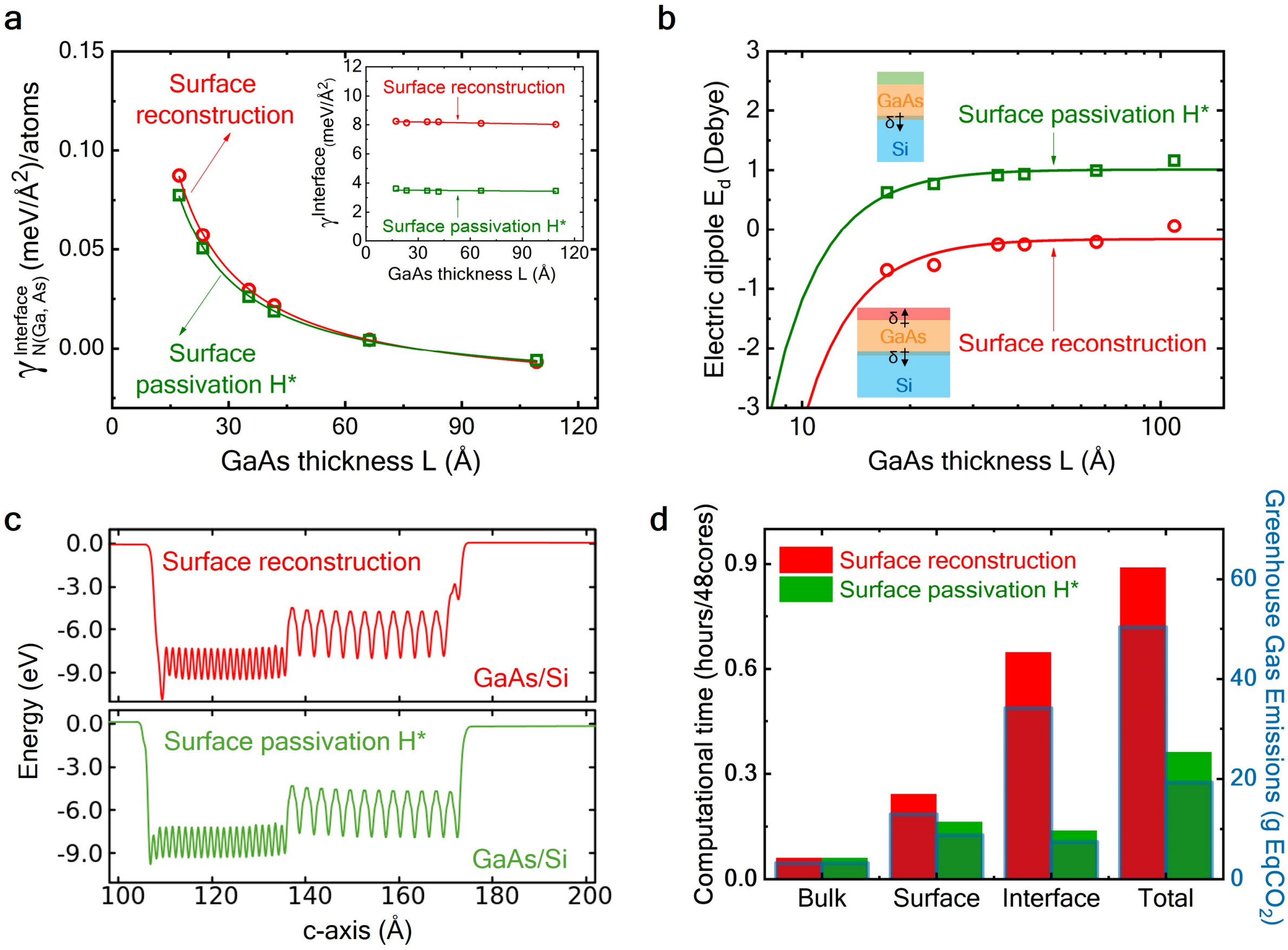

**Fig. 3. Impact and advantages of H* passivation methods.** **a** Bi-axially strained GaAs/Si(001) interface energy per atoms ($\gamma^{\text{Interface}}_{\text{N(Ga,As)}}$) with increasing GaAs thickness (L) under Ga-rich conditions (with solids lines as a guide-to-the-eye (see Supplementary note 2)); corrected interface energies are shown in the inset. **b** Total electric dipole values as a function of material thickness (L) for both methods, fitted with solids lines (see Supplementary note 2), dipole directions are illustrated in the inset figures. **c** Electrostatic potential diagram showing the Hartree potential as a function of the supercell thickness along the c-axis in Å (z-direction), comparing both approaches. **d** Comparison of the two methods in terms of computational time and greenhouse gas emissions.

### Impact on electric dipole interactions and electrostatic potentials

To further assess the potential of CLAPS method, we analyzed the evolution of the total electric dipole $E_d(L)$ (in Debye) as a function of the GaAs thickness considered on the substrate, using both the reconstructed surface and passivated surface technique, as shown in Fig. 3b, with the vacuum thickness fixed at 400 Å. The solid lines are the best fit using the analytical expressions for dipole-dipole interactions[62] for both curves. Details of the fitting parameters are provided in Supplementary note 2. Our previous studies demonstrated that at a vacuum spacing of at least 150 Å is necessary to minimize dipole interactions between periodic images, as evidenced by the highly polar GaP(111)A Ga-trimer surface slab[23]. In the abrupt interface case chosen in this work, H* passivation captures only the dipoles at the interface (pointing downward from Ga to Si as illustrated in Fig. 3b) while eliminating surface charge interactions. In contrast, the surface reconstruction method includes a combined dipole contribution from both the reconstructed surface and the interface, reflecting misaligned and mixed charge interactions (as illustrated in Fig. 3b). In the same Figure 3b, electric dipoles calculated using the H* passivated heterostructure are consistent with the analytical expression of dipole-dipole interactions (solid line), whereas the surface-reconstructed heterostructure exhibits deviations at lower GaAs thickness[62]. This discrepancy likely arises from additional dipole-dipole and multipole interactions, consistent with observations in other heterointerfaces[63,64]. The possible multipole interactions (Supplementary Fig. S5) influencing the surface and interface decrease with increasing GaAs thickness. This analysis sheds light on the influence of induced dipole interactions on the surface and interface, and their impact on absolute interface energy calculations. In this case, in addition to axial dipoles, the reconstructed surface method exhibits parasitic dipole effects induced by surface stress. In addition, a lateral periodicity size-dependence analysis was performed by examining GaP/Si supercells with different lateral dimensions (1x, 2x, and 3x). The results reveal size-independent interface energies and dipole moments, with the electric dipole scaling linearly with periodicity. This behavior demonstrates the robustness and consistency of our findings (see supplementary for details Fig. S6). Moreover, reconstructed surface method shows in-plane inhomogeneity in the spatial distribution of charges related to the surface reconstruction. The results indicate that the H* passivated heterostructure preserves bulk-like charge distribution at the surface, while the surface-reconstructed configuration exhibits pronounced disturbances and excess charge accumulation in the top layer, thus potentially impacting interface studies (Supplementary Fig. S7). We then investigated the effect of dipole interactions as a function of vacuum thickness. Using localized basis set approach, we were able to extend the vacuum region up to 10,000 Å, whereas plane-wave codes typically employ an average vacuum of 15-20 Å in their supercells. As shown in Fig. S8, our use of a 300–400 Å vacuum region is already sufficient to effectively eliminate dipole artifacts.

Furthermore, we analyzed the electrostatic potential profiles for both configurations and found that the reconstructed GaAs/Si interface (red) exhibits pronounced potential fluctuations near the surface–vacuum region, indicating charge accumulation and interfacial electric fields arising from unsaturated surface states (as shown in Fig. 3c). In contrast, the pseudo-H* passivated surface (green) displays a smooth and stable potential profile, confirming the suppression of charge accumulation and the enhancement of electrostatic uniformity across the interface. However, the biaxially strained interface induces an internal electric field within the slab, resulting in a shift in the potential profile.

## Impact on computational cost and greenhouse gas emissions

Moreover, we evaluated the impact of the proposed approach for calculating the absolute interface energy on both computational time and associated greenhouse gas ($CO_2$) emissions. The emissions were estimated following the GENCI methodology, which provides standardized emission factors for different computing architectures. For CPU-based calculations, an emission rate of 1.1 g Eq$CO_2$/core/hour was used to convert computational time into the corresponding carbon footprint. As shown in Fig. 3d, we quantified the computational time and resulting emissions for each step involved in determining the GaP/Si interface energy. Single-point calculations on optimized structures were carried out using 48 cores (one full node) of the Skylake partition at the Très Grand Centre de Calcul du CEA (TGCC) high-performance computing facility.

The results reveal substantial differences in computational cost between the surface reconstruction and pseudo-H* passivation approaches. The surface passivation method achieves over 60% reductions in both computational time and $CO_2$ emissions compared to the reconstruction method, highlighting its superior efficiency and lower environmental impact. It is also noteworthy that these single-point calculations represent only a small fraction of the total computational demand required for full relaxation calculations. This notable reduction comes from the smaller system size, reduced number of atoms, and lower charge requirements, all inherent to the use of CLAPS .

### Impact on electronic states: band structure analysis

Finally, we evaluated the impact of the proposed CLAPS on the band structure analysis of the III-V/Si heterostructure. Figure 4a, outlined by a red rectangle, presents the total band structure of the GaP/Si heterostructure obtained using the surface reconstruction approach, showing multiple views of the same band structure, each corresponding to projections onto different atoms while the underlying band structure remains unchanged. The band structure is inherently complex, as it contains contributions from two surfaces, two bulk regions, and the interface. To distinguish these contributions, the total band structure was projected onto atoms in the top surface, bottom surface, and interface using a weighted band analysis, also known as the fat-band method, as implemented in SIESTA.

At first, the top surface region was examined for both Ga and P atoms, showing partial electronic state projections near the Fermi level, indicative of defect-induced surface states. Next, the interface atoms (e.g. Ga and Si) were analyzed, revealing well-defined yet highly dispersed interface states located deep within the electronic bandgap, without any ambiguity. Additionally, strong defect-related contributions were observed from the bottom Si surface, where these states, although relatively non-dispersed compared to the interface states, are also located deep within the electronic bandgap. This suggests that both the top and bottom surface states collectively influence the apparent interface states near the Fermi level. These non-physical contributions emphasize the critical need to mitigate surface effects in band structure calculations to faithfully represent the intrinsic properties of the interface.

We then compared these results with the CLAPS, as shown in Fig. 4b (outlined by a light green rectangle), which shows the total band structure of the GaP/Si heterostructure using surface passivation H*. Most importantly, the non-physical contributions arising from surface artifacts have completely disappeared. Indeed, the projected electronic states of the top surface H* and Ga atoms show no significant contributions near the Fermi level, unlike the previous case with surface reconstruction. Additionally, the partial electronic state projections near the Fermi level, caused by the defect-induced surface states, are eliminated. Similarly, the projected interface states of the Ga and Si atoms at the abrupt interface exhibit pristine electronic states, free from surface-induced defect features. Now, the bottom-surface H atoms also show no discernible effect near the Fermi level. These results demonstrate that, unlike the surface reconstruction method, surface passivation effectively eliminates surface states near the Fermi level.

Notably, all calculations presented above were performed using the GGA exchange-correlation functional[65]. To confirm the validity of the results, hybrid functional corrections (e.g., HSE) were incorporated into the GGA calculations to ensure that electron localization and self-interaction errors remain negligible.  Specifically, the HSE06 hybrid functional was employed to analyze the band structure (see details in methods). HSE06 provides more accurate band structures and band offsets, while also reducing the artificial charge transfer that arises from GGA's underestimated band gap. Figure 4c (also outlined by a light green rectangle) presents the element-projected band structure of the GaP/Si heterostructure, where the DFT framework was advanced from the conventional GGA to the hybrid HSE06 level of theory. In the present case, this opens the band gap and slightly shifts the electronic bands near the Fermi level, improving the overall description of the electronic structure. Using this higher level of theory, the interface states are more clearly isolated and remain independent of other states, providing a more reliable description of the heterointerface electronic structure.

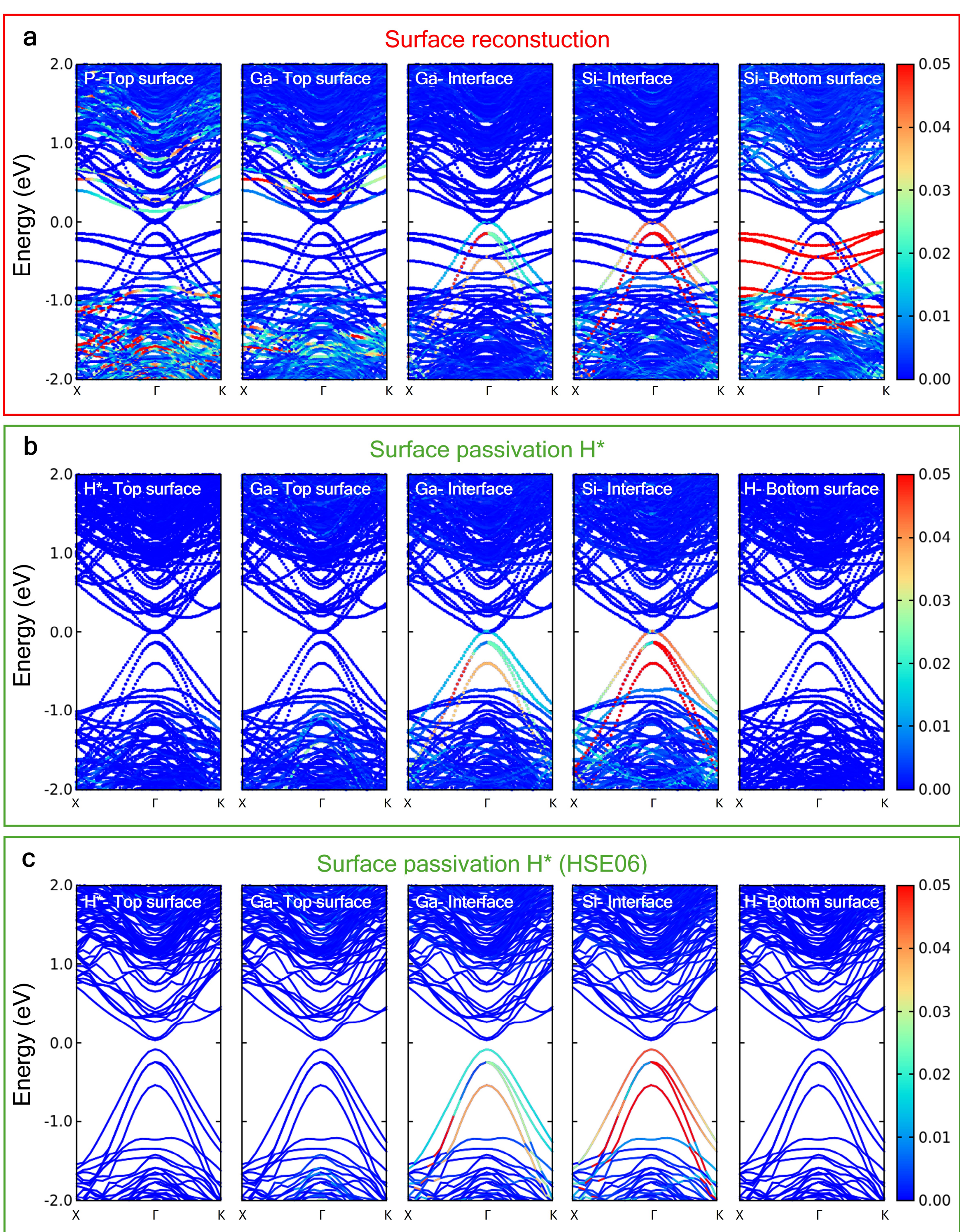

**Fig. 4. Electronic structure analysis of the GaP/Si heterostructure.** Band structure diagrams showing the projected atomic contributions from the top surface atoms, interface atoms, and bottom surface atoms for: (a) surface reconstruction (GGA), (b) surface passivation (GGA), and (c) surface passivation using the hybrid functional HSE06. The projection intensity is represented using a scaling factor of 0.05.

## Discussion

We have introduced above a computationally efficient approach for determining absolute interface energies in heterostructures using density functional theory with localized basis sets and surface passivation technique so-called CLAPS. Notably, this approach expresses the absolute interface energy as a function of chemical potential, clarifies interface energetics within the thermodynamic range, and employs the combination of surface passivation technique (here using pseudo-H*) on the surfaces and localized basis sets to systematically address and mitigate common dipolar sources of error. These errors include uncertainties arising from guessed reconstructions, finite-size effects, convergence issues, and dipole artifacts. The robustness and versatility of the CLAPS were demonstrated through its application to both quasi-lattice-matched (GaP/Si) and lattice-mismatched (GaAs/Si) heterostructures. Despite the use of high k-point calculations, the surface passivation technique enhances computational efficiency and significantly reduces resource consumption. Notably, it lowers greenhouse gas emissions by 60% compared to the surface reconstruction method. Furthermore, it yields a clearer and more reliable representation of the electronic states in the band structure, enabling more accurate analysis of heterostructure properties. In addition, CLAPS is computationally efficient, enabling streamlined heterointerface calculations and the *ab initio* determination of absolute interface energies. This approach systematically investigates the fundamental principles governing a wide variety of heterointerface configurations and is applicable to a broad class of material heterostructures (e.g. III-V, II-VI, IV semiconductors, oxides, perovskites, metals and semi-metals, 2D materials, molecular assemblies), including plastically relaxed systems. For interfaces involving highly ionic materials, correction schemes, such as varying the charge of the surface pseudo-atoms as proposed by M. Stengel *et al.*[66] and S.-H. Yoo *et al.*[67], may be employed. Overall, the approach provides an efficient framework and serves as a valuable tool for optimizing high-quality, cost-effective, and multifunctional devices across a broad range of applications, particularly in next-generation photonics, nanoelectronics, quantum technologies, energy conversion, and other emerging fields where precise interface control is essential.

## Methods

### Computational details:

All DFT calculations[52,53] were performed using the SIESTA code[48-50], employing a basis set of finite-range numerical pseudo-atomic orbitals for the valence wave functions. The exchange-correlation effects were treated within the Generalized Gradient Approximation (GGA) functional in the Perdew-Burke-Ernzerhof (PBE) form[65]. Core-valence interactions were described by Troullier-Martins (TM) norm-conserving pseudopotentials[68]. Brillouin zone integrations were performed using a 4x2x1 and 6×6×1 Monkhrost-Pack k-point grid for surface reconstructed and H* passivated interfaces studies, respectively[69]. All the structures were minimized with the force criteria of 0.005 eV/Å. For the strained GaAs/Si heterostructures, a real-space mesh cutoff of 300 Ry was employed, while a cutoff 150 Ry was used for the quasi-lattice-matched GaP/Si studies. To prevent interactions between periodic images, a vacuum region of approximatively 400 Å was introduced along the vertical (z) direction of the substrate. Pseudopotentials for the pseudo-Hydrogen-like (H*) atoms were generated using the ATOM code within the SIESTA package. The total energies vary linearly with system size, as shown in Fig. S8. HSE06 hybrid functional calculations were carried out using a 1×1×1 Monkhorst-Pack k-point grid with the HONPAS package, implemented within the SIESTA framework[70].

### GaAs: Heat of formation energy

In a materials system, the chemical potential must satisfy specific thermodynamic constraints. The maximum values of the chemical potentials of arsenic $\mu_{\mathrm{As}}$ and gallium $\mu_{\mathrm{Ga}}$ are achieved when each element exists in its own pure bulk phase, leading to the conditions: $\mu_{\mathrm{As}} < \mu_{\mathrm{As}}^{\mathrm{As-bulk}}$ and $\mu_{\mathrm{Ga}} < \mu_{\mathrm{Ga}}^{\mathrm{Ga-bulk}}$.

Furthermore, under thermodynamic equilibrium, the chemical potentials must satisfy the stability condition of the GaAs bulk phase: $\mu_{Ga} + \mu_{As} = \mu_{GaAs}^{GaAs-bulk}$. These constraints ensure that the GaAs material remains stable without decomposition into its elemental constituents. Thus,

$$\mu_{GaAs}^{GaAs-bulk} = \mu_{Ga}^{Ga-bulk} + \mu_{As}^{As-bulk} + \Delta H_f(GaAs) \quad (3)$$

Where $\Delta H_f(GaAs)$ denotes the heat of formation of GaAs material. In this work, $\Delta H_f(GaAs)$ is calculated to be -0.70 eV, which is in good agreement with the experimentally reported value -0.74 eV [56]. This energy defines the thermodynamic stability window between Ga-rich and As-rich growth conditions, as illustrated in Fig. 2b.

The GaAs/Si interface energy is evaluated as a function of the variation in the arsenic chemical potential, defined as $\Delta\mu_{As} = \mu_{As} - \mu_{As}^{As-bulk}$. The thermodynamic limits for $\Delta\mu_{As}$ are given by:

$$\Delta H_f(GaAs) < \Delta\mu_{As} < 0 \quad (4)$$

when $\Delta\mu_{As}$ equals the heat of formation $\Delta H_f(GaAs)$, the system reaches the Ga-rich extreme limit, favoring the formation of bulk Ga. In contrast, when $\Delta\mu_{As}$ equals 0, the As-rich limit is achieved, leading to the preferential formation of bulk As.

## Absolute interface energy using reconstructed surfaces

This approach enables the determination of all possible interface energies, independent of both the crystal orientation and the atomic structure of the interface. It also requires a detailed understanding of the free surface reconstructions and their corresponding energies. However, it involves heavy DFT calculations and substantial computational time. We define the absolute interface energy as ${}^{I}\gamma_{Bottom}^{Top}$, where $I$ denote the studied interface. Top and Bottom refer to the specific surfaces of the slab associated with the two materials under consideration. The absolute interface energy is defined as:

$${}^{I}\gamma_{Bottom}^{Top} = \frac{E_{slab}^{int} - \sum_i \left(N_i \mu_i^{i-bulk}\right) - \sum_{S=Top,Bottom}\left(A\gamma_{surf}^{S}\right)}{A} \quad (5)$$

where $E_{slab}^{int}$ is the total energy of the slab, $\mu_i^{i-bulk}$ denote the chemical potentials of atom i in its bulk phase, $N_i$ is the number of atom i, $\gamma_{surf}^{S}$ represents the surface energies of the top or the bottom surfaces, and A is the in-plane surface area of the slab. Interestingly, Eqn. (5)

highlights the explicit dependence of the interface energy on the chemical potentials of the constituent elements and the specific surface energies selected.

From the equation above, we can derive the relationship for the absolute interface energy of GaAs/Si, ${}^{I}\gamma_{Si}^{GaAs}$, where the GaAs(001)β2(2x4) surface is on top of the slab at the GaAs part and the reconstructed Si(001)c(4x2) surface is at the bottom of the slab at the Si part.

$$ {}^{I}\gamma_{Si}^{GaAs} = \frac{E_{slab}^{GaAs/Si} - N_{GaAs}\mu_{GaAs}^{GaAs-bulk} - (N_{Ga} - N_{As})\mu_{As} - N_{Si}\mu_{Si}^{Si-bulk} - A\gamma_{surf}^{GaAs} - A\gamma_{surf}^{Si}}{A} \quad (6) $$

where $E_{slab}^{GaAs/Si}$ is the total energy of the slab, $N_{Ga}$ and $N_{As}$ represent the number of Ga and As atoms in the slab, $\mu_{GaAs}^{GaAs-bulk}$ and $\mu_{As}$ are the chemical potentials of the GaAs bulk and As species, and A is the surface area. $\mu_{Si}^{Si-bulk}$ is the chemical potential of silicon bulk, $N_{si}$ is the number of Silicon atoms, and $\gamma_{surf}^{GaAs}$ and $\gamma_{surf}^{Si}$ are the specific surface energies per unit area for the top and bottom surfaces of GaAs and Si, respectively.

## Data availability

All data needed to evaluate the conclusions in the paper are present in the paper and/or the Supplementary Information. The other data that supports the findings of this study are available from the corresponding author upon request.

## Acknowledgements

This research was supported by the French National Research NUAGES Project (Grant no. ANR-21-CE24-0006). DFT calculations were performed at FOTON Institute, and the work was granted access to the HPC resources of TGCC/CINES under the allocation A0140911434, A0160911434, and A0180911434 made by GENCI.

## Author contributions

S.P.C. conducted the investigation, carried out formal analysis, curated the data, and prepared the original draft. S.A. contributed to data analysis and manuscript review and editing. C.C. and L.P. conceived the study, performed methodology development, supervised the project, validated the results, contributed to data analysis, manuscript review and editing, and contributed to project administration and conceptualization. C.C. acquired funding, and L.P. was responsible for resource management.

## Competing interest

The authors declare no competing interests.

## Additional information

**Supplementary information** The online version contains supplementary material available at

**Correspondence** and requests for materials should be addressed to Charles Cornet and Laurent Pedesseau.